\documentclass{iaus}
\usepackage{graphicx}

\title[Neutron Stars in Globular Clusters] 
{Neutron Stars in Globular Clusters}

\author[Ivanova, Heinke \& Rasio]   
{N. Ivanova$^1$, C. O. Heinke$^{2,3}$ \break \and F. Rasio$^3$}

\affiliation{$^1$
CITA, University of Toronto, 60 St George St, Toronto, ON M5R 2N6, Canada 
\break email: nata@cita.utoronto.ca\\[\affilskip]
$^2$Department of Astronomy, University of Virginia, 530 McCormick Road
Charlottesville, VA 22904-4325, USA \break email: cheinke@virginia.edu\\[\affilskip]
$^3$Physics and Astronomy Department, Northwestern University, 2145 Sheridan Rd, 
\break Evanston, IL 60208 USA \break email: rasio@northwestern.edu}

\pubyear{2008}
\volume{246}  
\pagerange{119--126}
\date{?? and in revised form ??}
\jname{Proceedings Title IAU Symposium}
\editors{A.C. Editor, B.D. Editor \& C.E. Editor, eds.}
\begin{document}

\maketitle

\begin{abstract}
Dynamical interactions that occur between objects in dense stellar systems 
are particularly important for the question of formation of X-ray binaries. 
We present results of numerical simulations of 70 globular clusters with
different dynamical properties and a total stellar mass of $2\times 10^7 M_\odot$.   We find that
in order to retain enough neutron stars to match observations we must assume that NSs can be formed via electron-capture supernovae.
Our simulations explain the observed dependence of the number of LMXBs on ``collision number'' as well as the large scatter observed between different globular clusters.  For millisecond pulsars, we obtain good agreement between our models and the numbers and characteristics of observed pulsars in the clusters Terzan 5 and 47 Tuc.
\end{abstract}

\firstsection 
\section{Introduction}

In globular clusters (GCs), neutron stars (NSs) are seen 
as low-mass X-ray binaries (LMXBs), bright or  quiescent,
and as binary or single millisecond pulsars (MSPs).
As the numbers per unit mass of LMXBs and MSPs in clusters greatly exceed their numbers in the Galaxy, their origin has been linked to stellar encounters that should occur frequently in an environment of high stellar density (Clark 1975).
Galactic GCs are known to contain thirteen bright X-ray sources, 
and as many as $\sim$100-200 quiescent LMXBs (qLMXBs) are thought to exist in the Galactic GC system (Heinke et al. 2003).
Millisecond pulsars, likely descendants of LMXBs (for a review see, e.g., Bhattacharya and van den Heuvel 1991),
also are present in GCs in great numbers:  
about 140 GC millisecond radio pulsars have been detected by 
now\footnote{See http://www.naic.edu/~pfreire/GCpsr.html for an updated list.},
with more than a dozen in several GCs -- in 47~Tuc (Camilo et al. 2000),
M28 (Stairs et al. 2006),  and Terzan~5 (Ransom et al. 2005). 
As only a few X-ray binaries or few dozen MSPs are present per fairly massive (more than a million stars) 
and dense cluster, computationally this problem is  very challenging.
The target time for direct $N$-body methods to address the million-body problem is around 2020 
(Hut 2006). In our studies, we use a modified encounter rate technique method, described in detail in 
Ivanova et al. (2005), and with the updates described in Ivanova et al. (2007).

\section{Production and retention}

In our studies we adopt that a NS can be formed as a result of either a core-collapse (CC) supernova 
or an electron capture supernova (ECS). In the latter case, three possibilities are considered:
ECS during normal stellar evolution, accretion induced collapse (AIC) and merger induced collapse.
For the case of CC NSs, we adopt that a supernova was accompanied by a natal kick in accordance 
with the distribution by Hobbs et al. (2005). For ECS NSs, we adopt that the accompanying natal kick is 
10 times smaller. We find that even considering a stellar population with 100\%  primordial binaries,
the retention fraction of CC NSs is very small (Fig.~1) and the resulting number of retained NSs is just a few per 
 typical dense globular cluster of $2\times10^5 \ M_\odot$.
NSs formed via different ECSs channels are retained in reasonable numbers, providing
about 200 retained NSs per  typical GC,
or more than a thousand in a cluster like 47 Tuc (similar numbers were found also in Kuranov \& Postnov, 2006).
Therefore, in contrast to the population of NSs
in the Galaxy, the population of NSs in GCs is mainly low-mass NSs made by ECS. 

\begin{figure}
\begin{center}
\includegraphics[height=3.5in,width=3.5in]{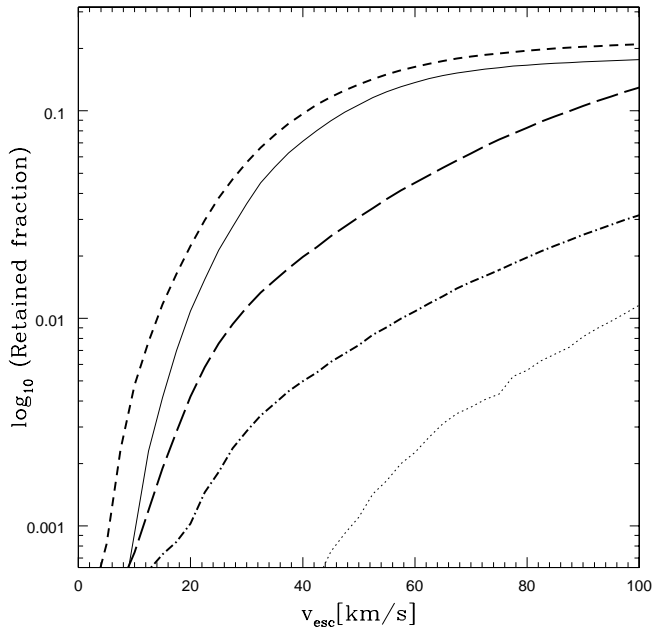}
  \caption{The retention fractions as a function of escape velocity (for stellar evolution 
unaffected by dynamics) for a Hobbs et al. (2005) kick distribution. 
Dotted and dash-dotted lines show the retention fractions for single and binary 
populations,
core-collapse NSs only. Solid and short-dashed lines show the total retention 
 fractions for single and binary populations, all NSs. 
For comparison, we show  the total retention fraction of a binary 
population with the Arzoumanian et al. (2002) kick distribution (long-dashed line).
}\label{fig:ret}
\end{center}
\end{figure}

\section{X-ray binaries}

In our simulations, we find that a typical GC can contain up to
2 LMXBs with a MS companion (most likely observed, at any particular time, as qLMXBs) and up to one LMXB with
a WD companion  (ultra-compact X-ray binaries, UCXBs). The scatter in the average number of observed LMXBs per cluster in  
independent simulations is very large - e.g., for UCXBs, it can vary between 0.1 and 1.1. 
In the case of Terzan 5 and 47 Tucanae, the average number of LMXBs formed per Gyr, at the 
age of 11 Gyr, is $\sim 5$ for NS-MS LMXBs and $\sim 8$ for UCXBs. 
These numbers are in general agreement with the observations.
Overall the numbers of NS that gain mass via mass transfer (MT) through 11 Gyr of cluster evolution
are high: for our 47 Tuc model, about 40 NS-MS binaries and more than 70 UCXBs.
As we observe fewer MSPs in these GCs, while the rate of LMXB formation in simulations 
is consistent with the observations, we conclude that not all NSs that gain mass via MT become currently active MSPs.

We analyzed how the specific number of LMXBs $n_{\rm LMXB}$ in our simulations depends on the specific collision frequency
(see Fig.~2). For the case when only core density is varied, $n_{\rm LMXB}$ depends linearly on $\gamma$.
Variation of other cluster dynamical properties leads to deviation from such a linear dependence, which may explain the scatter in $\gamma$ in the observed GCs.

\begin{figure}
\begin{center}
\includegraphics[height=3.5in,width=3.5in]{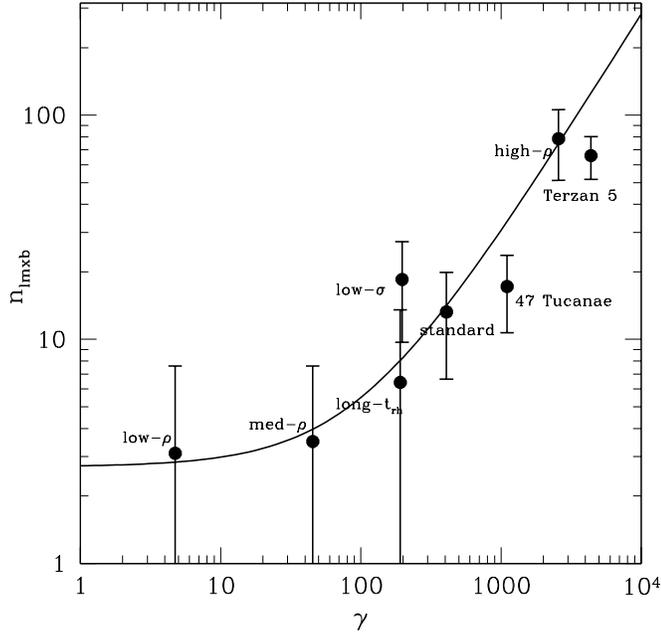}
  \caption{The collision numbers $\gamma$ (Verbunt \& Hut 1987) and numbers of LMXBs in simulated clusters.
The solid line corresponds to $n_{\rm LMXB} = (2.7\pm6) +0.028\gamma$, $\gamma$ is 
per $10^6 M_\odot$, as in Pooley \& Hut (2006).  
The error bars correspond to the scatter in our simulations.
The standard model has a central core density of $10^5$ pc$^{-3}$, velocity dispersion 10 km/s
and half-mass relaxation time 1 Gyr. 
The low-$\rho$, med-$\rho$, standard and high-$\rho$ models have central core densities
$10^3$, $10^4$, $10^5$ and $10^6$ pc$^{-3}$, respectively. Our low-$\sigma$ model has velocity 
dispersion 5 km/s, while long-$t_{\rm rh}$ has half-mass relaxation time 3 Gyr.
}\label{fig:gam}
\end{center}
\end{figure}

\section{Millisecond pulsars}

\begin{figure}
\begin{center}
\includegraphics[height=2.5in,width=2.5in]{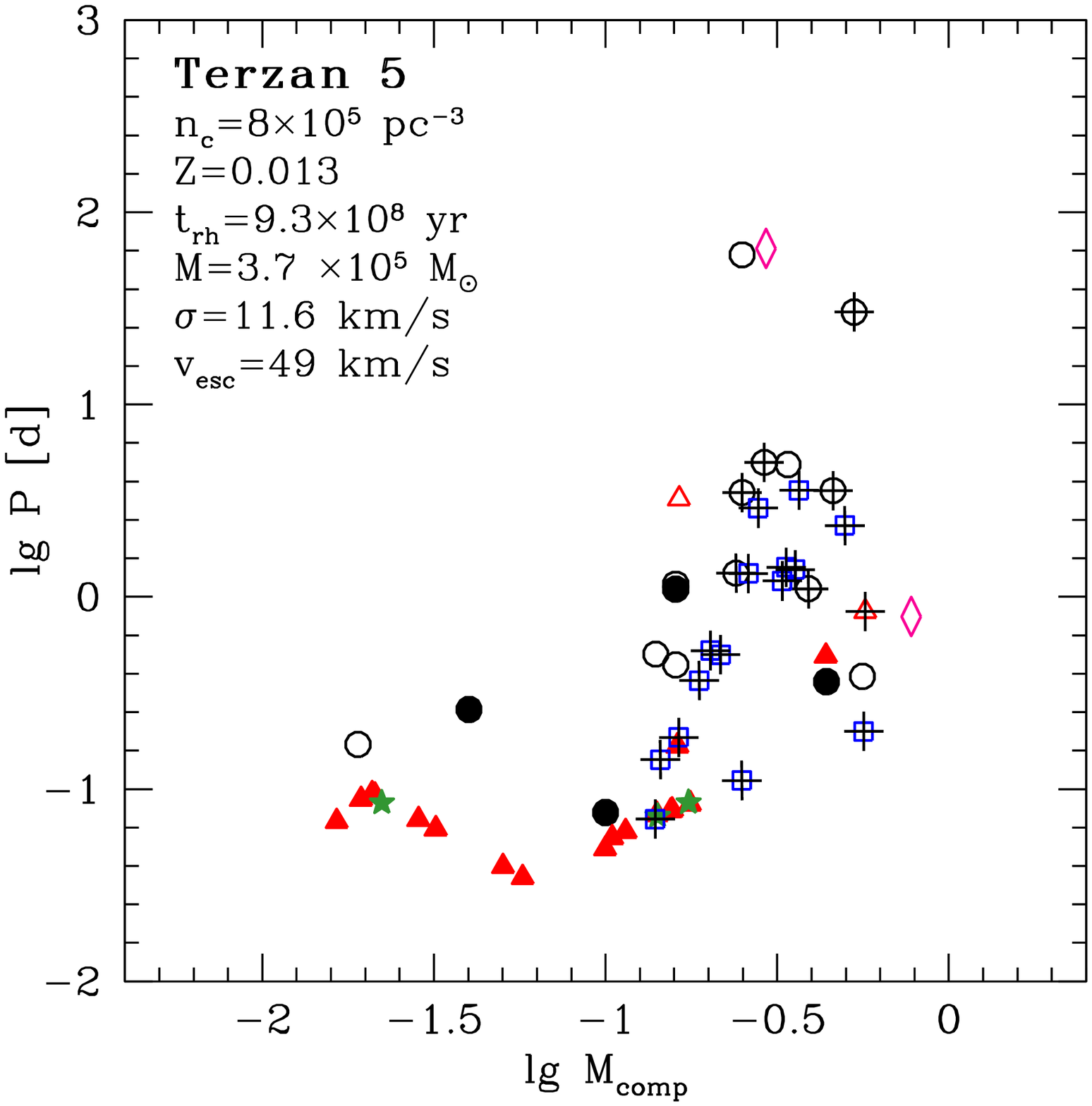}
\includegraphics[height=2.5in,width=2.5in]{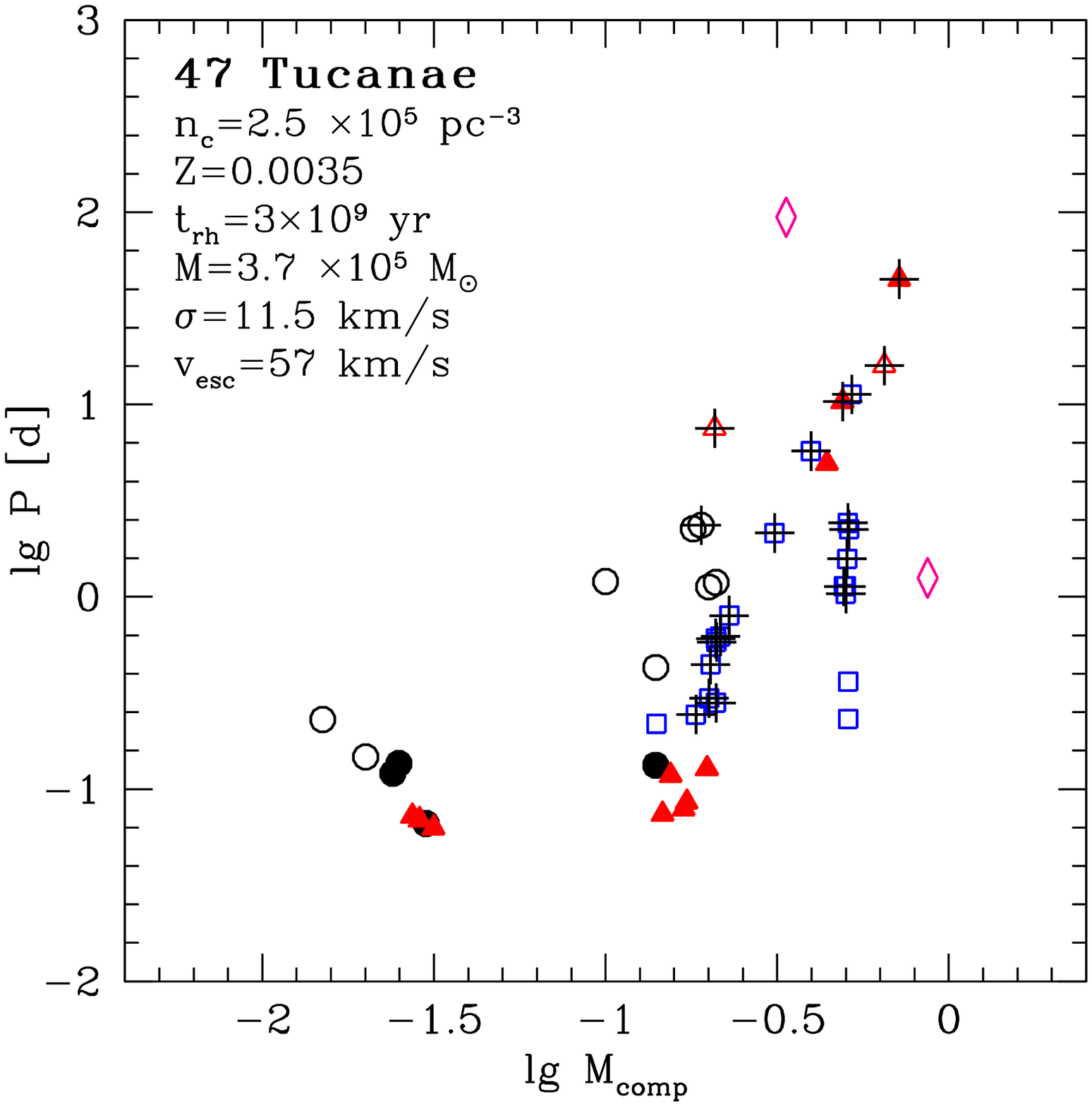}
  \caption{bMSPs in simulated models of 47 Tuc and Terzan 5 compared to observed bMSPs.  The simulated populations correspond to several independent runs and represent a larger 
population than in the observed clusters. Observed bMSPs are shown with circles;
triangles - bMSP formed via binary exchanges; stars - via tidal captures;
squares - via physical collisions; diamond - primordial binaries. 
 Cross signs mark eccentric bMSPs ($e\ge0.05$) and solid symbols mark systems with a non-degenerate companion (in the case of simulations) or observed eclipsing systems.  
}\label{fig:msp}
\end{center}
\end{figure}

Suppose that all mass-gaining events in the life of a NS -- mass transfer,
physical collision with a red giant, common envelope hyper-accretion or merger --
can lead to NS recycling.
In this case we found that as many as 250 and 320 potential MSPs
are made in our simulations of clusters like Terzan 5 or 47 Tuc, accordingly 
(the corresponding numbers of retained NSs are $\sim 500$ and $\sim 1100$).
Although these numbers correlate well with the formation rate of LMXBs, 
it greatly exceeds the numbers of observed and inferred MSPs in both clusters, 
which are 33 (in Terzan 5; perhaps 60 total) and 22 (in 47 Tucanae; perhaps 30 total).

We analyzed the population of NSs that gained mass.
We found that bMSPs formed from primordial binaries, 
where a common envelope event led to AIC,
create a population of potential bMSPs with relatively heavy companions, in circular orbits with periods from one day to several hundreds of days. 
This population is not seen in either Terzan 5 or 47 Tuc. 
We considered primordial binaries that evolved through  
mass transfer from a giant donor after a NS was formed via AIC.  
Even though bMSPs that have similar periods, companion masses and eccentricities
are present in Terzan 5, there are no such systems in 47 Tucanae.
Also, bMSPs made from primordial binaries after AIC must inevitably be formed in low-dense clusters, but no bMSPs are observed there. 
These facts tell us that either AIC does not work, or the kicks in the case of AIC are stronger then we adopted, 
or a NS formed via AIC has such a strong magnetic field, that surface accretion does not occur.

A common understanding of MSP formation is that the NS is recycled through disk 
accretion, where a NS is spun up only if the accretion rate is not too low, $\dot M \ge 3\times 10^{-3}\dot M_{\rm Edd}$,
where $ \dot M_{\rm Edd}$ is the Eddington limit (for a review see, e.g., Lamb \& Yu 2005).
In a UCXB, soon after the start of mass transfer,
the accretion rate drops very quickly.  After 1 Gyr,
it is less than $10^{-4} \dot M_{\rm Edd}$. Such a MT leads to a spin-down of the previously spun-up NS, and no MSP is formed. 
Support for this statement is given by the fact that no 
UCXBs (those that have WD companions) are visible as MSPs (Lamb \& Yu 2005).

The requirement of steady spin-up through disk accretion implies that not all physical collisions
will lead necessarily to NS spin-up. In the case of a physical collision with a giant,
the NS will retain a fraction of the giant envelope, with a mass of a few hundredths of $M_\odot$ (Lombardi et al. 2006).
Immediately after the collision, this material has angular momentum and most likely will form a disk. 
We adopted therefore that in the case of a physical collision with a giant, an MSP can be formed, but, 
in the case of any other physical collision, the NS will not be recycled.

Considering all the exclusions described above, we form in our simulations at least   
$15\pm7$ MSPs for Terzan 5 and $25\pm4$ MSPs for 47 Tuc (for the formed population of bMSPs, see Fig.~3).  
The values for Terzan 5 are somewhat uncertain due to uncertainty in the properties of this heavily reddened cluster.
The total number of NSs that gain mass in the simulations by one or other way are 250 and 320
in Terzan 5 or 47 Tuc, accordingly.

\section{Conclusions}

We studied the formation and evolution of NSs in globular clusters. We find that
NS formation via different channels of ECS is very effective in GCs
and provides most of the retained NSs.
Having as many as a few hundred retained NSs per typical GC,
or about 1100 per cluster like 47 Tuc, produces LMXBs in numbers comparable 
to observations. We note that if AIC does not lead to the formation of NSs,
then the number of formed NSs is reduced only by $\sim 20\%$, but the number
of appearing LMXBs is decreased by 2-3 times (per Gyr, at the cluster age of 11 Gyr),
although it may still be consistent with the observations, given the large
scatter in the simulations.  
We find that up to half of NSs could gain mass after their formation through
 mass transfer, hyper-accretion during a common envelope, or physical collision. 
It is likely that most of these mass-gaining events do not lead to  NS spin-up, 
and that only a few per cent of all NSs appear eventually as MSPs, implying that there is a large underlying population of unseen NSs in GCs.

\end{document}